\documentclass[letterpaper,english]{spie}

\usepackage[T1]{fontenc}
\usepackage[latin9]{inputenc}
\usepackage{babel}
\usepackage{textcomp}
\usepackage{amsmath}
\usepackage{amssymb}
\usepackage{graphicx}
\usepackage[numbers]{natbib}
\usepackage[unicode=true,pdfusetitle,
 bookmarks=true,bookmarksnumbered=false,bookmarksopen=false,
 breaklinks=false,pdfborder={0 0 1},backref=false,colorlinks=false]
 {hyperref}

\makeatletter

\pdfpageheight\paperheight
\pdfpagewidth\paperwidth

\@ifundefined{showcaptionsetup}{}{%
 \PassOptionsToPackage{caption=false}{subfig}}
\usepackage{subfig}
\makeatother

\begin{document}

\title{Notes on Evanescent Wave Bragg-Reflection Waveguides}

\author{Benedikt Pressl and Gregor Weihs\skiplinehalf Institut für Experimentalphysik,
Universität Innsbruck,\\
 Technikerstraße 25, 6020 Innsbruck, Austria}
\maketitle
\begin{abstract}
We investigate an extended version of the Bragg reflection waveguide
(BRW) with air gaps as one of the layers. This design has the potential
of drastically simplifying the epitaxial structure for integrated
nonlinear optical elements at the expense of more complicated structuring.
This approach would afford much more flexibility for designing and
varying BRW structures. Here, we discuss an extension of the established
theory for BRW slabs and report our results of applying Marcatili's
method for rectangular waveguides to the BRW case. With this analytic
approach we can estimate the effective index of the modes orders of
magnitudes faster than with full numerical techniques, such as finite-difference
time-domain (FDTD) or finite elements. Initial results are mixed;
while phase-matched designs have been found, they currently have no
significant advantage over other schemes.
\end{abstract}


\authorinfo{B.P.: Email: benedikt.pressl@uibk.ac.at}

\section{Introduction}

Nonlinear optical elements are interesting building blocks for a wide
range applications, ranging from coherent sources for broadband spectroscopy
via classical frequency conversion processes to the generation of
entangled photon pairs with parametric down conversion. For practical
purposes, miniaturization and integration on semiconductor platforms
seems imperative - with the (aluminium) gallium arsenide (AlGaAs)
platform as a possible choice. These materials have extraordinarily
high nonlinear coefficients and refractive indices, which increase
efficiency and decrease device size, respectively. However, owing
to the cubic lattice structure, no kind of natural birefringence exists
in GaAs, which is commonly exploited for phase matching. Thus, quasi-phase-matching
or modal phase matching has to be employed. In the case of Bragg-reflection
waveguides (BRW) \citep{Yeh1976}, modal phase matching between the
fundamental mode and the second harmonic mode is achieved with distributed
Bragg reflectors enclosing a core \citep{Abolghasem2011,Horn2012}.
In a simple picture one can assume that the fundamental mode is confined
by total internal reflection of the core, while the second harmonic
mode is defined by the reflection of the distributed Bragg reflector
(DBR) stacks. This allows different (almost independent) tuning of
the modes \citep{West2006}.

\begin{figure}
\begin{centering}
\subfloat[\label{fig:2DEWBRW}]{\centering{}\includegraphics[width=0.45\textwidth]{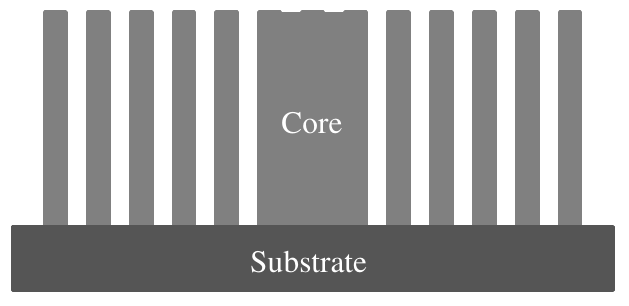}}\hspace*{\fill}\subfloat[\label{fig:Index-profile-of1DEWBRW}]{\begin{centering}
\includegraphics[width=0.45\textwidth]{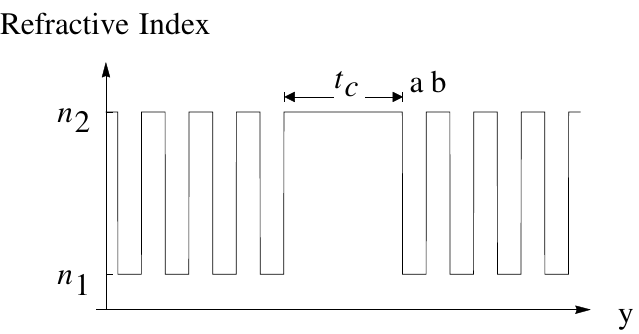}
\par\end{centering}

}
\par\end{centering}

\caption{\label{fig:Definitions-for-the}In our design (\ref{fig:2DEWBRW})
of an evanescent wave BRW the horizontal (etched) structure forms
the Bragg reflector whereas the vertical structure is a conventional
dielectric waveguide. The definitions for the 1D simplification (\ref{fig:Index-profile-of1DEWBRW})
of the same waveguide are used in combination with Marcatili's method. }
\end{figure}

In this paper we investigate a different - more generalized - way
of designing the distributed Bragg reflector DBR stacks. Usually,
the DBR structure is grown vertically by alternating layers of AlGaAs
with different aluminum content and thus different refractive indices.
This means that it is fixed for an entire wafer, which may conflict
with other structures and devices one may want to integrate on the
chip. Horizontal confinement is achieved by total internal reflection
at the sidewalls of etched ridges. In this article in contrast, we
propose a simple two-layer wafer, where the Bragg structure is achieved
by etching thin trenches (figure (\ref{fig:Definitions-for-the}))
on the sides of a core region. Here, vertical confinement is provided
by total internal reflection while the horizontal confinement is provided
by the DBR stack. Such a structure would afford significantly more
flexibility in accommodating nonlinear interaction with other behavior
or functionality. For example, the DBR could be introduced adiabatically
with the goal of transferring the mode of a relatively large waveguide
into the desired mode in the DBR region. Electrical current paths
for integrated pump lasers would not have to traverse the relatively
thick DBR. Finally, a single epitaxially grown wafer could yield many
substantially different devices thus enabling a shorter design cycle
and a wider range of adaptation to varying wavelengths or other characteristics
of the desired optical processes.

Since in this approach one of the Bragg layers is simply air (or some
kind of filling material with a low refractive index) the field becomes
evanescent in these layers. Similar structures have already been studied
in different contexts: in general as photonic crystals \citep{Joannopoulos2008},
as waveguide grating reflectors \citep{Chen2006} or more specifically
in photon tunneling experiments \citep{PhysRevE.64.026609}.

The analytical theory of BRWs is well established\citep{Yeh1976,West2006,Abolghasem2011}.
However, it fails if any of the material refractive indices are smaller
than the effective index of the mode involved. The evanescent field
components require special care as several simplifying assumptions
break down. In the following we will first lay out the fundamentals
of the evanescent BRW and then show how Marcatili's method can be
applied for a very efficient solution of the problem at hand. Finally
we will draw our conclusions as to the usefulness of the concept regarding
practical devices.

\section{Slab Waveguide Effective Index Calculation for the Evanescent Case}

In order to calculate the effective index of a mode, Maxwell's equations
have to be solved. Bragg reflection waveguides possess symmetries
that allow to derive an exact analytical expression for the modes
\citep{Yeh1976} if only one dimension is considered and the Bragg
stack is assumed to be infinitely extended. 

The basic formalism is already described in \citep{Yariv2006} and
follows \citep{West2006}. We carried out calculations starting with
Maxwell's equations to verify that there are no hidden assumptions
that invalidate the treatment of evanescent fields. These evanescent
fields mean that the wavevector is complex valued in certain parts
of the structure. This prohibits an easy analytical solution for the
dispersion relation; it has to be found numerically. Nevertheless,
using analytic formulations as far as practical reduces the numerical
workload to just a couple of equations instead of dealing with huge
matrices, such as in the case of a gridded, full-vectorial, 2D eigenmode
calculation.

In general, the eigenvalue equation (\ref{eq:TE-Dispersion-Equation})
determines the dispersion of a mode specified by the boundary conditions
between the core wavefunction $f_{c}$ (left hand side) and the wavefunction
in the Bragg stack (right hand side). This (standard) approach yields
two equations with two unknown factors for the continuity of the function
and the derivative, respectively. Dividing these equations eliminates
these constants and results in the following form:

\begin{equation}
\frac{1}{k_{c}}f_{c}(k_{c}t_{c}/2)\left[\left.\frac{\partial f_{c}}{\partial x}\right|_{k_{c}t_{c}/2}\right]^{-1}=\frac{i}{k_{1}}\frac{\exp\left[iK\Lambda\right]-A+B}{\exp\left[iK\Lambda\right]-A-B}\label{eq:TE-Dispersion-Equation}
\end{equation}

In the case of a simple step index core $f_{c}$ takes the form of
either $\sin$ or $\cos$, depending on whether we are interested
in symmetric or antisymmetric modes. The left hand side then reduces
to a $\tan$ or $\cot$ function; using the form of equation (\ref{eq:TE-Dispersion-Equation})
assures proper factors in the case of higher order modes. $t_{c}$
is the core thickness and $k_{i}$ is the wavevector in layer $i$
(see also eq. (\ref{eq:k-plug-in})). The right hand side derives
from the eigenvalues of a single Bragg stack. $A,B$ (transfer matrix
entries) and $\exp\left[iK\Lambda\right]$ (Bloch-wave eigenvalues)
are properties of the Bragg stack, with

\begin{align}
\exp\left[iK\Lambda\right]= & \frac{1}{2}(A+D)\pm\sqrt{\frac{1}{4}(A+D)^{2}-1}\label{eq:Eigenvalue}\\
A_{\mathrm{TE}}= & e^{ik_{1}a}\,\left[\cos k_{2}b+\frac{i}{2}\left(\frac{k_{2}}{k_{1}}+\frac{k_{2}}{k_{1}}\right)\sin k_{2}b\right]\\
B_{\mathrm{TE}}= & e^{-ik_{1}a}\left[\frac{i}{2}\left(\frac{k_{2}}{k_{1}}-\frac{k_{2}}{k_{1}}\right)\sin k_{2}b\right]
\end{align}

being the coefficients for a TE mode (for the TM analogue see Refs.
\citep{Yariv2006,West2006}). $a$ and $b$ are the thickness of the
two layers of the Bragg stacks with $\Lambda=a+b$. The wavevectors
in the individual layers (with vacuum wavevector $k_{0}=2\pi/\lambda$)
are given by

\begin{equation}
k_{i}=k_{0}\sqrt{n_{i}^{2}-n_{\mathrm{eff}}^{2}}\label{eq:k-plug-in}
\end{equation}

With equations (\ref{eq:Eigenvalue}) to (\ref{eq:k-plug-in}) the
dispersion relation (\ref{eq:TE-Dispersion-Equation}) is now solely
a function of the effective refractive index of the mode $n_{\mathrm{eff}}$.
Additional assumptions, such as $n_{i}>n_{\mathrm{eff}}$, lead to
an analytical expression for $n_{\mathrm{eff}}$ \citep{West2006}.
In general, however, one of the $k_{i}$ can be complex, so we have
to find numerical solutions in the complex domain ($n_{\mathrm{eff}}\in\mathbb{C}$).
This is a two dimensional root-finding problem, but since we are only
interested in low-loss solutions we can effectively restrict the search
to the region where $\mathrm{Im\,}n_{\mathrm{eff}}\approx0$. Additionally,
both branches of (\ref{eq:Eigenvalue}) have to be taken into account
for finding the proper solution.

\begin{figure}
\begin{centering}
\subfloat[\label{fig:Anti-symmetric-core-modes}Anti-symmetric core modes (sine)]{\begin{centering}
\includegraphics[width=0.45\textwidth]{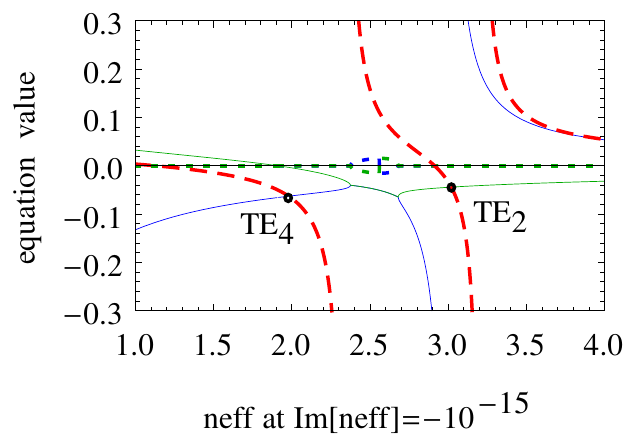}
\par\end{centering}

}\hspace*{\fill}\subfloat[\label{fig:Symmetric-(cosine)-core}Symmetric (cosine) core modes,
the fundamental mode is at $n_{\mathrm{eff}}\approx3.23$ ]{\begin{centering}
\includegraphics[width=0.45\textwidth]{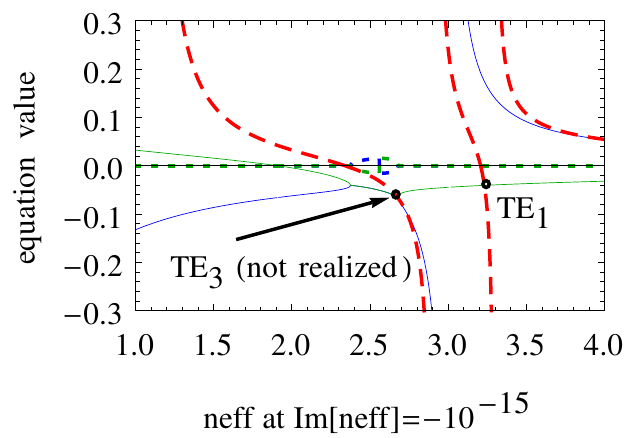}
\par\end{centering}

}
\par\end{centering}

\caption{\label{fig:LBRW1} Graphical solutions of (\ref{eq:k-plug-in}) for
the refractive indices $n_{1}=1$, $n_{2}=n_{c}=3.3$, core thickness
$t_{c}=0.5\,\mathrm{\text{\textmu}m}$, layer thicknesses $a=b=0.1\,\mathrm{\text{\textmu m}}$
at a wavelength $\lambda=0.775\,\mathrm{\text{\textmu m}}$. The solid
lines represent the real part of the right hand side of the equation
(Bragg stack) while the dashed lines are the left hand side (core
function). There are two branches for the Bragg stack (blue and green)
that intersect with the core function (dashed, red) at $n_{\mathrm{eff}}\approx2$
($\mathrm{TE_{4}}$), $n_{\mathrm{eff}}\approx3$ ($\mathrm{TE_{2}}$)
and $n_{\mathrm{eff}}\approx3.2$ ($\mathrm{TE_{1}}$), thus defining
three modes. At $n_{\mathrm{eff}}\approx2.5$ the real parts of both
branches join and form a ``cutoff'' region as the imaginary parts
(dotted) depart from the axis. Despite the intersection of the core
function with the real parts at $n_{\mathrm{eff}}\approx2.65$ (corresponding
to $\mathrm{TE_{3}}$) in figure (\ref{fig:Symmetric-(cosine)-core}),
no complex solution can be found here and so the mode is not realized. }
\end{figure}

Figure (\ref{fig:LBRW1}) shows the graphical solutions for the refractive
indices $n_{1}=1$, $n_{2}=n_{c}=3.3$, core thickness $t_{c}=0.5\,\mathrm{\text{\textmu}m}$,
layer thicknesses $a=b=0.1\,\mathrm{\text{\textmu m}}$ at a wavelength
$\lambda=0.775\,\mathrm{\text{\textmu m}}$. There are two physical
solutions for the effective mode index at $n_{\mathrm{eff}}=3.0196-9\cdot10^{-14}i$
($\mathrm{TE_{2}}$ mode, with two peaks) and $n_{\mathrm{eff}}=1.979-2\cdot10^{-14}i$
($\mathrm{TE_{4}}$ mode). In comparison, a commercial-grade mode
solver \citep{LumericalSolutions} yields $n_{\mathrm{eff}}=3.0192-9\cdot10^{-16}i$
and $n_{\mathrm{eff}}=1.991+9\cdot10^{-17}i$, showing excellent agreement.
The inconsistent signs of the imaginary parts are due to limited machine
precision.

\section{Marcatili Method for (Evanescent Wave) Bragg Reflection Waveguides}

Having solved the slab waveguide, Marcatili's method \citep{Marcatili1969}
for estimating the modes of rectangular waveguides can directly be
employed for the full 2D problem. In this approximation it is assumed
that both directions (e.g. $x$ and $y$ for a $z$-propagating mode)
are independent slab waveguides. Both slab waveguides (here, 1D-BRW
for $x$; asymmetric, dielectric waveguide for $y$) are solved and
then are joined by the approximate dispersion relation (see \citep{Chen2006})

\begin{equation}
\beta^{2}=\beta_{x}^{2}+\beta_{y}^{2}-\beta_{c}^{2}
\end{equation}

with $\beta=n_{\mathrm{eff}}k_{0}$, the desired propagation constant.
$\beta_{x}$ and $\beta_{y}$ are the propagation constants of the
respective slabs ($\beta_{i}=n_{\mathrm{eff,i}}k_{0}$), $\beta_{c}=n_{c}k_{0}$
is the propagation constant of a plane wave in a bulk of core material.

Using the previous example and the structure from figure (\ref{fig:2DEWBRW}),
we introduce the slab height $h$, the substrate refractive index
$n_{S}$ and the ``background index'' $n_{bg}=1$ (air). In such
a configuration the substrate plays an important role: if the effective
index of a mode cannot be less than the refractive index of the substrate.
Here, a substrate index of $n_{S}=3$ and the vertical confinement
- which further reduces the effective index - is already enough to
push the $\mathrm{TE_{2}}$ found in the 1D case into the cutoff region
(figure (\ref{fig:Anti-symmetric-core-modes})). As an example, Marcatili's
2D method predicts an effective index of $n_{\mathrm{eff}}=3.211+9\cdot10^{-13}i$
for the $\mathrm{TE_{1}}$ mode for an evanescent wave BRW with a
height of $h=1\,\mathrm{\text{\textmu m}}$, while the commercial
grade mode solver gives $n_{\mathrm{eff}}=3.202+10^{-17}i$. In this
case, a larger deviation than for the slab is expected as Marcatili's
method commonly underestimates the propagation index \citep{Chen2006}.

\section{Discussion and Summary}

One advantage of these semi-analytical models is computational speed
as only a few equations have to be solved. This allows many more structures
to be analyzed for feasibility than with ``numerically exact'' mode
solvers. Results for the intended structures etched structures (such
as (\ref{fig:2DEWBRW})) are mixed: for the 1-D slab waveguide, configurations
that phase-match a fundamental mode at $\lambda=1.55\,\mathrm{\text{\textmu m}}$
with a second harmonic (Bragg) mode at $\lambda=0.775\,\mathrm{\text{\textmu m}}$
can be found easily. With the added vertical confinement of a realistic
2D structure, however, it is much more difficult as the substrate
imposes a very restrictive limit on the possible mode indices. At
present state there is no significant advantage in terms of tunability,
easiness in fabrication or nonlinear overlap compared to, for example,
form-birefringence phase-matched nano-waveguides \citep{Rutkowska2011}.
Some of these limitations are related to the extreme refractive index
difference between GaAs and air and the relatively high substrate
refractive index. Having a filling material - for example, a polymer
\citep{Reitzenstein2011} - for the trenches with a refractive index
of > 2 would relax the requirements.
\begin{acknowledgments}
This work was supported in part by the Quantum Information Processing
Program of the Canadian Institut for Advanced Research (CIFAR) and
by the European Research Council, Project 257531 ``EnSeNa''.
\end{acknowledgments}
\bibliographystyle{spiebib}
\bibliography{lbrw}

\begin{thebibliography}{10}

\bibitem{Yeh1976}
P.~Yeh and A.~Yariv, ``{Bragg Reflection Waveguides},'' {\em Optics
  Communications}~{\bf 19}(3), pp.~427--430, 1976.

\bibitem{Abolghasem2011}
P.~Abolghasem, {\em Phase-Matching Second-Order Optical Nonlinear Interactions
  using Bragg Reflection Waveguides: A Platform for Integrated Parametric
  Devices}.
\newblock PhD thesis, University of Toronto, 2011.

\bibitem{Horn2012}
R.~Horn, P.~Abolghasem, B.~J. Bijlani, D.~Kang, A.~S. Helmy, and G.~Weihs,
  ``{Monolithic Source of Photon Pairs},'' {\em Phys. Rev. Lett.}~{\bf 108},
  p.~153605, 2012.

\bibitem{West2006}
B.~R. West and A.~S. Helmy, ``{Properties of the quarter-wave Bragg reflection
  waveguide: theory},'' {\em J. Opt. Soc. Am. B}~{\bf 23}(6), 2006.

\bibitem{Joannopoulos2008}
J.~Joannopoulos, S.~Johnson, J.~Winn, and R.~Meade, {\em Photonic Crystals:
  Molding the Flow of Light (Second Edition)}, Princeton University Press,
  2008.

\bibitem{Chen2006}
C.-L. Chen, {\em Foundations for guided-wave optics}, Wiley, 2006.

\bibitem{PhysRevE.64.026609}
S.~Esposito, ``Universal photonic tunneling time,'' {\em Phys. Rev. E}~{\bf
  64}, p.~026609, 2001.

\bibitem{Yariv2006}
A.~Yariv and P.~Yeh, {\em {Photonics: Optical Electronics in Modern
  Communications}}, The Oxford Series in Electrical and Computer Engineering,
  Oxford University Press, USA, 2006.

\bibitem{LumericalSolutions}
{Lumerical Solutions, Inc.}

\bibitem{Marcatili1969}
E.~A.~J. Marcatili, ``Dielectric rectangular waveguide and directional coupler
  for integrated optics,'' {\em Bell Sys. Tech. J.}~{\bf 48}, pp.~2071--2102,
  1969.

\bibitem{Rutkowska2011}
K.~A. Rutkowska, D.~Duchesne, M.~Volatier, R.~Ar\`es, V.~Ameiz, and
  R.~Morandotti, ``{Second Harmonic Generation in AlGaAs Nanowaveguides},''
  {\em Acty Physica Polonica A}~{\bf 120}(4), pp.~725--731, 2011.

\bibitem{Reitzenstein2011}
S.~Reitzenstein, T.~Heindel, C.~Kistner, F.~Albert, T.~Braun, C.~Hopfmann,
  P.~Mrowinski, M.~Lermer, C.~Schneider, S.~H\"ofling, M.~Kamp, and A.~Forchel,
  ``{Electrically Driven Quantum Dot Micropillar Light Sources},'' {\em IEEE
  Sel. Top. Quant. Electron.}~{\bf 17}(6), pp.~1670--1680, 2011.

\end{thebibliography}

\end{document}